\documentclass[12pt]{iopart}
\usepackage{epsf}
\begin{document}

\title{Rashba spin-orbit coupling and spin 
precession in carbon nanotubes}

\author{A De Martino and R Egger\footnote[3]{To whom 
correspondence should be addressed (egger@thphy.uni-duesseldorf.de)} }

\address{Institut f{\"u}r Theoretische Physik, 
Heinrich-Heine-Universit{\"a}t, D-40225 D{\"u}sseldorf}

\begin{abstract}
The Rashba spin-orbit coupling in carbon nanotubes and its 
effect on  spin-dependent transport properties are 
analyzed theoretically.  
We focus on clean non-interacting nanotubes with tunable 
number of subbands $N$. 
The peculiar band structure is shown to allow in principle for 
Datta-Das oscillatory behavior in the tunneling magnetoresistance as a 
function of gate voltage, despite the presence of multiple bands.
We discuss the conditions for observing Datta-Das oscillations in
carbon nanotubes.
\end{abstract}


\submitto{\JPCM}

\maketitle

\section{Introduction}

Spintronics in molecular conductors is a field attracting more
and more attention, both from fundamental physics as well as
from application-oriented material science \cite{zutic}.  
Here the quantum-mechanical electronic spin is the central
object controlling transport properties. For a conductor sandwiched
between ferromagnetic leads, a different resistance can be observed
depending on the relative orientation of the lead magnetizations.
Quite often, the resistance is larger in the antiparallel configuration
than in the parallel one, but sometimes also the reverse situation can
be observed. It is useful to define the tunnel magnetoresistance
(TMR), $\rho_t= (R_{AP}-R_P)/R_P$, as the relative difference between
the corresponding resistances.  

A particularly interesting material in that context is provided
by carbon nanotubes (CNTs), see Refs.~\cite{forro,ando05} for
general reviews.
Quite a number of experimental studies concerning spin transport
through individual multi- (MWNT) or single-walled (SWNT) nanotubes
contacted by ferromagnetic leads have been reported over the past few years
\cite{alphenaar,schneider,kim,chakra,hoffer,kontos,sahoo}.
In particular, the experiments of the Basel group 
\cite{kontos,sahoo} use
thin-film PdNi alloys as ferromagnetic leads in order
to contact either SWNTs or MWNTs,
where the shape anisotropy and the geometry of the setup 
allow for the study of the spin-dependence of electrical transport. These
experiments have revealed
{\sl oscillatory} behavior of the TMR as a function of the external
gate voltage.  Similar oscillations were predicted as a consequence
of the gate-voltage-tunable Rashba spin-orbit (SO) interaction 
\cite{rashba,bychkov}
in a classic paper by Datta and Das some time ago \cite{datta}.  
Since Datta-Das oscillations have still not been observed experimentally
so far, a thorough theoretical investigation of this effect
in nanotubes
is called for and provided here. 
Unfortunately, from our analysis below, we find that the weakness of
SO couplings in nanotubes excludes an 
interpretation of these data in terms of the Datta-Das effect 
-- they can, however, be explained 
in terms of quantum interference effects
\cite{sahoo}.  Nevertheless, we show that the
presence of multiple bands in CNTs is not detrimental,
and under certain circumstances, the effect may be sufficiently enhanced
to be observable, e.g., by a tuning of the number of bands via external
gates along the lines of Ref.~\cite{strunk}.
In the original Datta-Das proposal \cite{datta}, subband mixing was
ignored so that different channels just add up coherently, but subband
mixing has later been argued to spoil the effect \cite{mireles,valin}. In CNTs,
the special band structure requires a careful re-examination of 
the Datta-Das idea in this context, and we shall show that the
arguments of Refs.~\cite{mireles,valin} do not necessarily apply here.

Recent theoretical studies of spin-dependent transport in CNTs
have mainly focused on the single-channel limit,
taking into account electron-electron interactions  within
the framework of the Luttinger liquid theory
 \cite{balents,ando,ourprl,jpcm,hausler2}
(see also \cite{hausler1,zulicke,gritsev}
for related discussions on interacting quantum wires 
with Rashba SO coupling).
Here we confine ourselves to the noninteracting problem in
order to not overly complicate the analysis, but study
the many-band case and details of the band structure.  
Interactions can be taken into account 
within the Luttinger liquid approach at a later stage,
and may enhance the effect of SO
couplings \cite{hausler1,chenraikh}.  
We shall also neglect disorder
effects.  Mean free paths in high-quality SWNTs typically 
exceed $1 \mu$m, while in MWNTs this may be a more severe
approximation for some samples.  However, high-quality MWNTs with
ultra-long mean free paths have also been reported recently 
\cite{urbina}. 

The structure of this paper is as follows. In Sec.~\ref{sec2} we derive
the Rashba spin-orbit hamiltonian from microscopic considerations. 
The resulting tight-binding SO
hamiltonian will be studied at low
energy scales in Sec.~\ref{sec3}, where we derive  its
continuum form.  In Sec.~\ref{sec4}, the consequences
with regard to Datta-Das oscillations in the TMR are analyzed.
We shall always consider the zero-temperature limit, and (in
 most of the paper) put $\hbar=1$.

\section{Rashba spin-orbit coupling in nanotubes}\label{sec2}

We start by noting that transport effectively proceeds through the outermost
shell of a MWNT only, such that we can take a single-shell model even
when dealing with a MWNT.
Experimentally and theoretically, it is understood
that such a model works very well in good-quality MWNTs \cite{forro},
essentially because only the outermost shell is electrically
contacted and tunneling between different shells is largely
suppressed \cite{bachtold,christian}.
Naturally, a single-shell description is also appropriate for SWNTs,
where we assume a sufficiently large radius $R$ such that occupation
of multiple subbands can be possible.  (For a MWNT, $R$ denotes the
radius of the outermost shell.)
Depending on the electrochemical potential $\mu$ (doping level), 
we then have to deal with $N$ spin-degenerate bands.
We assume full quantum coherence (no dephasing), so that the usual
Landauer-B\"uttiker approach applies, and
exclude external magnetic fields or electric field inhomogeneities, say,
due to the electrodes.  We proceed to derive the Rashba
SO interaction, $H_{so}$, for this problem. 
Notice that this is different from the intrinsic 
atomic SO interaction discussed in Refs.~\cite{ando,chico}.
In particular, the SO coupling in Refs.~\cite{ando,chico}
vanishes in the limit of large radius, which is not the case
for the Rashba SO coupling we discuss below. Though Ando's SO 
coupling \cite{ando} could straightforwardly be included in our analysis, 
being gate-voltage independent it could not change our conclusions 
relative to the gate-voltage dependent oscillations in the 
magnetoresistance and is neglected in what follows.

We first define a fixed reference frame 
${\cal S}= \{ \hat Y, \hat Z, \hat X \}$, with unit vector
$\hat X$ pointing in the axis direction and $\hat Z$ perpendicular to
the substrate on which the CNT is supposed to be located.
Next we introduce a second, local 
reference frame ${\cal S}_i = \{ \hat \rho_i, \hat t_i,\hat X\} $
relative to 
each lattice site 
$\vec R_i$ on the tube surface, where $\hat \rho_i$
and $\hat t_i$ are unit vectors along the local normal and tangential 
(around the circumference) directions at $\vec R_i$, respectively.
Using polar coordinates in the plane transverse to the tube axis,
the relation between ${\cal S}$ and ${\cal S}_i$ is given by 
\begin{equation}
\hat \rho_i = \cos \varphi_i \hat Y + \sin \varphi_i \hat Z , \quad
\hat t_i = -\sin \varphi_i \hat Y + \cos \varphi_i \hat Z .
\end{equation}
The position vector of a given carbon atom can then be written as 
$\vec R_i = R\hat \rho_i + X_i \hat X$.
For later convenience, we introduce also 
another reference frame. 
For each pair of sites $\vec  R_i$ and $\vec R_j$, we define
\begin{equation} \label{rij}
\vec R_{ij}  =\vec R_i -\vec R_j \equiv X_{ij} \hat X + 
\vec \rho_{ij},
\end{equation}
and denote the direction perpendicular to 
$\hat \rho_{ij}$ and $\hat X$ as $\hat \rho^\perp_{ij}$. Then 
$\{\hat \rho^\perp_{ij}, \hat \rho_{ij},\hat X \}$ constitutes a new local 
frame ${\cal S}_{ij}$, and one has
\begin{eqnarray}
\hat \rho^\perp_{ij} &=& \cos [(\varphi_i+\varphi_j)/2] \, \hat Y +
\sin [(\varphi_i+\varphi_j)/2] \, \hat Z ,\\ \nonumber
\hat \rho_{ij} &=& -\sin [(\varphi_i+\varphi_j)/2] \, \hat Y +
\cos [(\varphi_i+\varphi_j)/2] \, \hat Z . 
\end{eqnarray}
The $2p_z$ orbital at position $\vec R_i$ can then be represented as
\begin{equation}\label{2pz}
\chi_i(\vec r- \vec R_i) = \alpha
(\vec r -\vec R_i)\cdot \hat \rho_i
e^{-\beta |\vec r -\vec R_i|} ,
\end{equation}
where $4\alpha =(2\pi a_0^5)^{-1/2}$, 
$\beta =(2a_0)^{-1}$,  $a_0=\hbar^2/me^2=0.53 $\AA~is the Bohr radius,
and $m$ is the electron's mass.
We introduce an index $i$ on the orbital in order
to keep track of the atom at which it is centered.
The wavefunction (\ref{2pz}) is expected to be highly accurate 
for not too small $R$, where hybridization with the $sp^2$ 
orbitals is negligible.

At large distances from the tube,  external gates 
generally produce an electric field perpendicular 
to the tube axis and the substrate. 
As it has been shown in detail in previous works \cite{louie,saslow}, 
polarization effects of the CNT itself
due to a transverse field result in
a reduction of the externally applied 
field described by
\[
E_0 = \frac{1}{1+2\alpha_{0yy}/R^2} E_{ext},
\]
where $\alpha_{0yy}$ is the unscreened transverse static
polarizability.
Since $\alpha_{0yy}$ is approximately proportional to $R^2$, 
the factor in front of $E_{ext}$
practically equals a constant,
$\approx 0.2$ \cite{louie}.
Then, assuming homogeneity, the electric field due to 
the gate can be written as
\begin{equation}\label{e0}
\vec {E}= E_0 \hat Z,
\end{equation}
which in turn produces the (first-quantized) 
Rashba spin-orbit interaction \cite{rashba,bychkov}. With
standard Pauli matrices $\vec \sigma$ acting in spin space,
\begin{equation}\label{rashbac}
H_{so}= \frac{e\hbar}{4m^2c^2} \vec E \cdot ( {\vec\sigma} \times  \vec p ) .
\end{equation}

We proceed to derive the second-quantized spin-orbit hamiltonian
within the tight-binding approximation. For that purpose, we 
need the matrix element of the momentum operator between two 
$2p_z$ orbitals $\vec p_{ij} = \langle \chi_i |\vec p | \chi_j \rangle $,
from which we get
the following form for the SO hamiltonian:
\begin{equation}\label{so1}
H_{so} = g\sum_{ij} c_i^\dagger \left[
( \vec \sigma \times \vec p_{ij} ) \cdot \hat Z \right]
c_j ,
\end{equation}
where the fermionic operator $c_{i\sigma}$
destroys an electron with spin $\sigma=\uparrow,\downarrow$ 
in the $2p_z$ orbital centered at $\vec R_i$, and  
$g =E_0/4m^2c^2$.
For calculational convenience,
the matrix element $\vec p_{ij}$ can be written as 
$g \vec p_{ij} = i(\vec v_{ij}+\vec u_{ij})$,
where the spin-orbit vectors $\vec v_{ij}$ and $\vec u_{ij}$ 
are defined as
\begin{eqnarray}
\label{sovector1}
\vec v_{ij}  &=& -g \alpha
\int d^3\vec r \; \chi_i(\vec r -\vec R_{i})  
 \hat \rho_j e^{-\beta |\vec r -\vec R_j|} ,
\\ \label{sovector2}
 \vec u_{ij} &=& g \beta
\int d^3\vec r \; \chi_i(\vec r -\vec R_{i}) 
\frac{\vec r -\vec R_j}{|\vec r -\vec R_j |} \chi_j (\vec r- \vec R_j) ,
\end{eqnarray}
Note that the modulus
of $\vec v_{ij}$ and $\vec u_{ij}$
has dimension of energy, and 
their sum (but not necessarily each term separately) 
is antisymmetric under exchange of $i$ and $j$.

We first observe that the spin-orbit vectors connecting 
a site with itself clearly vanish, since
$\langle \chi_i |\vec p| \chi_i \rangle =0$.
Let us then discuss spin-orbit vectors connecting different sites. 
Since the orbitals (\ref{2pz}) decay exponentially, it is sufficient to
consider only the case of nearest neighbors. We start with $\vec v_{ij}$. 
Shifting $\vec r \rightarrow \vec s + \vec R_i$ in Eq.~(\ref{sovector1}) 
and using Eq.~(\ref{2pz}), we obtain
\[
\vec v_{ij} = - g \alpha^2
\hat \rho_j
\int d^3\vec s \; (\vec s\cdot \hat \rho_i ) e^{-\beta s}
e^{-\beta |\vec s +\vec R_{ij}|} .
\]
Using $\vec s= s_\parallel \hat R_{ij} + 
\vec s_\perp $, we then
rewrite the above integral as  
\[
\int d^3\vec s \; (s_\parallel \hat R_{ij} +
\vec s_\perp ) \cdot \hat \rho_i \, e^{-\beta s}
e^{-\beta \sqrt{(s_\parallel + d)^2 + s_\perp^2}} ,
\]
where we use $|\vec R_{ij}|=d$, with the nearest-neighbor distance
among carbon atoms in graphene $d=1.42$~\AA.
Note that $\beta d= 1.34$.
The second term in the brackets is odd in $\vec s_\perp$ and thus vanishes,
and we obtain
\begin{equation}
\vec v_{ij} = - g \alpha^2 \hat \rho_j \frac{2R}{d} \sin^2 
(\frac{\varphi_i-\varphi_j}{2}) d^4 \gamma_0 ,
\end{equation}
where we have used 
$\hat R_{ij}  \cdot \hat \rho_i = \frac{2R}{d}
\sin^2(\frac{\varphi_i-\varphi_j}{2} )$
and the dimensionless numerical factor $\gamma_0$ :
\[
\gamma_0 = \int dx\,dy\,dz \, x\,e^{-\beta d \sqrt{x^2+y^2+z^2}}
e^{-\beta d \sqrt{(x+d)^2+y^2+z^2}} .
\]
For $\vec v_{ji}$, we find
\[
\vec v_{ji} = g \alpha^2 \hat \rho_i \frac{2R}{d} \sin^2 
(\frac{\varphi_i-\varphi_j}{2}) d^4 \gamma_0 .
\]
Notice that, up to higher orders in $d/R$, the unit 
vectors $\hat \rho_{i,j}$ 
can be replaced by $\hat \rho^{\perp}_{ij}$, which makes clear that
$\vec v_{ij}$ is normal to the tube surface.
Now $|\sin[(\varphi_i-\varphi_j)/2]|$ varies between zero
(when the two sites are aligned in the axis direction) and 
$d/2R\ll 1$ (when the two sites are aligned in the circumferential
direction). 
Thus, to zeroth order in $d/R$, $\vec v_{ij}$ vanishes: 
it is a pure curvature effect, peculiar of nanotubes,
which does not exist in graphene. In practice,  $\vec v_{ij}$
is tiny and certainly subleading to $\vec u_{ij}$, which turns
out to be of order $(d/R)^0$. We shall therefore neglect it 
in what follows.

Let us now turn to $\vec u_{ij}$. We shift
$\vec r \to \vec s +(\vec R_i+\vec R_j)/2$ in Eq.~(\ref{sovector2}), 
and rewrite $\vec u_{ij}$ as the sum of two terms:
\begin{eqnarray}\label{int1}
\vec u^{(1)}_{ij} &=& g\beta \int
d^3\vec s \, \chi_i(\vec s -\vec R_{ij}/2) \,\chi_j(\vec s + \vec R_{ij}/2) 
\frac{\vec s}{|\vec s + \vec R_{ij}/2|} , \\
\label{int2}
\vec u^{(2)}_{ij} &=& \frac{g \beta}{2}
\vec R_{ij} \int
d^3\vec s \, \chi_i(\vec s -\vec R_{ij}/2) \, \chi_j(\vec s + \vec R_{ij}/2) 
\frac{1}{|\vec s + \vec R_{ij}/2|} .
\end{eqnarray}
Writing again $\vec s= s_\parallel \hat R_{ij} + \vec s_\perp$,
the computation of the above integrals leads, 
to the lowest non-vanishing order in $d/R$, to the following
expressions: 
\begin{eqnarray}\label{int3}
\vec u^{(1)}_{ij} &=& g\beta \alpha^2 \vec R_{ij} 
d^4 \gamma_1 \equiv u_1 \vec R_{ij} ,
\\
\label{int4}
\vec u^{(2)}_{ij} &=& \frac{g \beta}{2} \alpha^2 
\vec R_{ij} d^4 \gamma_2 \equiv u_2 \vec R_{ij} ,
\end{eqnarray}
with the dimensionless numerical factors
\begin{eqnarray} \label{gamma1}
\gamma_1 & = & \int dx\, dy\,dz 
\frac{x z^2 e^{-\beta d\sqrt{(x-1/2)^2+y^2+z^2}}
 e^{-\beta d\sqrt{(x+1/2)^2+y^2+z^2}}}
{\sqrt{(x+1/2)^2+y^2+z^2}} \\ \nonumber &\simeq&-0.0375,
\end{eqnarray}
and
\begin{eqnarray} \label{gamma2}
\gamma_2 &=& \int dx\, dy\,dz \frac{ z^2 e^{-\beta d \sqrt{(x-1/2)^2+y^2+z^2}}
 e^{-\beta d\sqrt{(x+1/2)^2+y^2+z^2}}} {\sqrt{(x+1/2)^2+y^2+z^2}} \\
&\simeq & 0.3748. \nonumber
\end{eqnarray}
To lowest order in $d/R$, it does not make a difference
whether we take the tangent unit vector
at $\vec R_j$, $\vec R_i$, or at $(\vec R_i+\vec R_j)/2$.
Hence we may write
$\hat \rho_{ij} \to \hat e_\varphi$, where $\hat e_\varphi$
is the unit tangent vector at $(\vec R_i +R_j)/2$. 
We then get SO couplings along the axial and along the
circumferential direction,
\begin{equation}\label{su}
\vec u_{ij} = u \left[ (\vec R_{ij} \cdot \hat X) \hat X +
 (\vec R_{ij} \cdot \hat e_\varphi) \hat e_\varphi \right] ,
\end{equation}
with $u=u_1+u_2$.  Note that
we have neglected a tiny component
of $\vec R_{ij}$ normal to the tube surface.
The above discussion then results in the tight-binding 
hamiltonian $H = H_0 + H_{so}$, where
\[
H_0 = - t \sum_{\vec r, a}  c^\dagger_{B,\vec r+\vec \delta_a } 
c^{}_{A,\vec r} + h.c., 
\]
with $t\approx 2.7$~eV \cite{ando05}.  Here the $\vec r$ denote all
sublattice-A tight-binding sites of the lattice. Furthermore,
the $\vec \delta_{a=1,2,3}$ are vectors connecting
$\vec r$ with the three nearest-neighbor sites which are all located
on sublattice B \cite{ando05}.
Since we consider the limit $d/R\ll 1$, 
the $\vec \delta_a$ at each site effectively lie in the 
tangent plane to the tube surface at that site.
The Rashba spin-orbit hamiltonian then  reads
\begin{equation}
\label{Hso} 
H_{so} = i u \sum_{\vec r, a} c^\dagger_{B,\vec r+\vec \delta_a } 
\Bigl[ \left( 
\vec \sigma \times 
[ (\vec \delta_a \cdot \hat X) \hat X +
  (\vec \delta_a \cdot \hat e_\varphi )  
\hat e_\varphi ] \right) 
\cdot \hat Z  \Bigr] 
c^{}_{A,\vec r} +h.c. 
\nonumber
\end{equation}

\section{Continuum limit}\label{sec3}

Since we are interested in the low-energy long-wavelength properties,
we now expand the electron operator around
the Fermi points $K,K'$ in terms of Bloch waves \cite{ando05},
\begin{equation}
\frac{c_{p\vec r}}{\sqrt{S}} = e^{i\vec K \cdot \vec r} F_{1p}(\vec r) +
e^{- i\vec K \cdot \vec r} F_{2p}(\vec r) , 
\end{equation}
where $S= \sqrt{3}a^2/2$ is the area of the unit cell, $a=\sqrt{3} d$,
and $p=A/B$ is the sublattice index. The $F_{\alpha p}$ are 
slowly varying electron field operators, and we choose the Fermi points at 
$\vec K = (4\pi/3a,0)$ and $\vec K' =  -\vec K$ \cite{ando05}. We 
then expand $F(\vec r + \vec \delta ) \simeq F(\vec r)
+ \vec \delta \cdot \nabla F(\vec r)$ and use the bond vectors 
\begin{equation} \label{delta}
\vec \delta_1 = \frac{a}{\sqrt{3}}(0,-1), \quad\quad
\vec \delta_2 = \frac{a}{2}(1, 1/\sqrt{3}), \quad\quad
\vec \delta_3 = \frac{a}{2}(-1,1/\sqrt{3}).
\end{equation}
These vectors are given in a fixed reference frame for a 2D graphene
sheet, and
we then must perform a rotation to longitudinal 
and circumferential directions via the chiral angle. This rotation
results in fixed phases that can be absorbed in the definition of 
$F_{\alpha p}$ and do not appear in final results. This is of course
expected from the $U(1)$ symmetry emerging at low energies in the dispersion 
relation of graphene \cite{ando05}.
After some algebra, the usual Dirac hamiltonian for the kinetic term follows,
\begin{equation}
H_0 = v \int \! d^2\vec r \; F^\dagger \left[ 
(T_0\otimes\tau_2\otimes\sigma_0)(-i\partial_x) +
(T_3\otimes \tau_1\otimes\sigma_0)(-i\partial_y)  
\right] F ,
\end{equation}
where $v=\sqrt{3}at/2\simeq 8\times 10^5$~m/sec is the Fermi velocity, 
 and $x,y$ are longitudinal and circumferential coordinates, 
respectively, with $0<y\leq 2\pi R$.  Finally, 
$T_i$ and $\tau_i$ are also Pauli matrices that now act in
the space of Fermi ($K,K'$) points and sublattice space ($A,B$), respectively.
For $i=0$, these are defined as $2\times 2$ unit matrices.

The low-energy limit of the SO term (\ref{Hso}) can be obtained in the 
following way. First we observe that
\[
\left[ (\vec \delta_a \cdot \hat X) 
\hat X  + 
(\vec \delta_a \cdot \hat e_\varphi ) \hat e_\varphi \right] \times  \hat Z =
- (\vec \delta_a\cdot \hat X) \hat Y  
- \sin (y/R) (\vec \delta_a \cdot \hat e_\varphi ) \hat X .
\]
Here the only approximation is the assumption that the bond vectors lie in 
the plane tangent to the nanotube surface at $\vec r$.
Second, by using the bond vectors (\ref{delta}) and taking into account 
the chiral angle $\eta$ between the fixed direction on the 
graphite sheet and the circumferential direction on the nanotube, 
one obtains
\begin{eqnarray}
\sum_a c^\dagger_{\vec r + \vec \delta_a} 
\left( \vec \delta_a \cdot \hat X  \right)
c_{\vec r}
& \approx & \frac{-3d}{2} \left( 
F^\dagger_{B1} e^{-i\eta}
F_{A1} +
F^\dagger_{B2} e^{i\eta} 
F_{A2} \right) , \nonumber \\
\sum_a c^\dagger_{\vec r + \vec \delta_a} 
\left( \vec \delta_a \cdot \hat e_\varphi  \right)
c_{\vec r}
& \approx & \frac{-3d}{2} 
\left( i F^\dagger_{B1} e^{-i\eta} F_{A1} 
 -i F^\dagger_{B2} e^{i\eta} F_{A2} \right) . \nonumber
\end{eqnarray}
Notice that we take into account exactly the relative orientation
of the bond vectors with respect to the directions $\hat X$ and 
$\hat e_\varphi$ for a generic nanotube, which is encoded in the
chiral angle $\eta$.
The constant phases $e^{\pm i \eta}$ can be absorbed by appropriately
redefining the operators as $F_{A2}\rightarrow e^{-i\eta } F_{A2}$ and
$F_{B1}\rightarrow e^{-i\eta } F_{B1}$,
and the final result can be written down in the form 
\begin{equation}\label{hsso}
H_{so} = \int \! d^2\vec r \; F^\dagger \left[ 
u_\parallel T_0\otimes\tau_1\otimes \sigma_2 
+ u_\perp \sin(y/R) T_3\otimes\tau_2 \otimes \sigma_1 \right] F ,
\end{equation}
with $u_{\parallel}=u_\perp = 3 d u/2$. 
For the sake of generality, we continue to use different 
coupling constants $u_\perp$ and $u_\parallel$.
It is worthwhile to mention that the leading term for the
Rashba spin-orbit coupling in a CNT, Eq.~(\ref{hsso}),
does not depend on longitudinal momentum.
This is due to the peculiar band structure of graphene with its isolated 
Fermi ($K$) points. In the above derivation, we also find terms 
that are linear in momentum, i.e., contain spatial 
derivatives of the electron operators.
Such terms  only produce 
tiny renormalizations of the velocities and
will be  neglected here.
The second term in Eq.~(\ref{hsso}) allows for spin flips and 
mixes transverse subbands.

{}From now on, for simplicity,  we consider just a single 
Fermi point, say, $K$. After the global $SU(2)$ rotation
 $\sigma_1\to \sigma_2 \to \sigma_3$ in spin space,
we get in compact notation
\begin{eqnarray} \label{unperham}
{\cal H}_0 &=&  v \left[ -i \tau_1 \partial_y - i \tau_2 \partial_x \right] . 
\\ \label{soham} 
{\cal H}_{so} &=&
u_\parallel \tau_1 \sigma_3 + u_\perp \sin(y/R)  \tau_2 \sigma_2 .
\end{eqnarray}
Note that the exact spectrum of ${\cal H}_0 + {\cal H}_{so}$ with $u_\perp=0$
can be obtained straightforwardly. In general, however, 
due to the smallness of the SO coupling (see below), it is enough to treat 
${\cal H}_{so}$ perturbatively. 
The following detailed derivation is then necessary 
to correctly evaluate the effect of the SO coupling, and moreover
it is interesting and important for the generalization to the 
interacting case, and for the analysis of features involving the 
electron wavefunction (as for instance electron-phonon interactions).

The eigenvalues of ${\cal H}_0$ are given by
\begin{equation} \label{eigzero} 
\epsilon_{an\sigma}(q) = av \sqrt{k^2_\perp(n) +q^2}\equiv a\epsilon_n(q)  ,
\end{equation} 
where $k_\perp(n) = (n+n_0)/R$ denotes the transverse momentum,
$q$ the longitudinal one, $a=\pm$ labels the conduction/valence band, 
and $\sigma=\pm$ the spin.
Here $n_0=0$ for intrinsically metallic shells, but
generally it can be taken as $0\leq n_0 \leq 1/2$ to take into account
chirality gaps or orbital magnetic fields along $\hat X$.  
The transverse subbands are labeled by integer values
$n=0,1,2,\ldots,{\cal N}-1$, where
${\cal N}= 2(N^2+M^2+NM)/{\rm gcd}(2M+N,2N+M)$ for ($N,M$) tubes
 \cite{ando05}.
${\cal N}$ is typically much larger than the actual number
$N=[k_F R]$ of occupied subbands, where we define $k_F=\mu/v$ with 
the doping level $\mu$ that we assume positive here.
The velocity $v_n$ for electrons in subband $n$ at the Fermi level
(in the absence of $H_{so}$)
and the corresponding Fermi momentum $q_n$ are then given by
\begin{equation}\label{vnqn}
v_n=v\sqrt{1-[(n+n_0)/(k_F R)]^2},\quad
q_n=k_F v_n/v. 
\end{equation}

The eigenvalues (\ref{eigzero}) are spin-independent
and thus doubly degenerate. 
The corresponding eigenstates are denoted  $| n q a \sigma \rangle$,
where $|n\rangle$ and $|q \rangle$ are respectively plane waves
in circumferential and longitudinal direction.
In coordinate representation they read 
\begin{equation}
\psi_{nqa\sigma}(x,y)\equiv \langle x,y| n q a \sigma \rangle =
\frac{e^{ik_\perp(n)y}}{\sqrt{2\pi R}} e^{iqx} \xi_{na}(q) \otimes
\chi_\sigma ,
\end{equation}
with the bispinor (in sublattice space)
\begin{equation}\label{bispinor2}
\xi_{n, a=\pm}(q) = \frac{1}{\sqrt{2}} \left(\begin{array}{c} 
e^{i \theta_n(q)/2} \\
\pm e^{- i\theta_n(q)/2} \end{array} \right) ,  \quad
e^{i \theta_n(q)} =  \frac{v (k_\perp (n) - i q ) }{ \epsilon_n(q)} .
\end{equation}
A different, and here more convenient basis is given by the sublattice 
states $| n q p  \sigma \rangle$.  Their coordinate representation is
\begin{equation}
\psi_{nqp\sigma}(x,y) =
\frac{e^{ik_\perp(n)y}}{\sqrt{2\pi R}} e^{iqx} \xi_p \otimes \chi_\sigma ,
\end{equation}
where $p=A,B$ and 
\[
\xi_A = \left( \begin{array}{c} 1 \\ 0 \end{array} \right) , \quad
\xi_B = \left( \begin{array}{c} 0 \\ 1 \end{array} \right) .
\]
Their usefulness stems from the fact that the $| n q p \sigma \rangle$ 
can be factorized as
\begin{equation}\label{factor}
| n q p \sigma \rangle = |n\rangle |q \rangle \otimes | p \sigma \rangle,
\quad | p \sigma \rangle = \xi_p \otimes \chi_\sigma,
\end{equation}
where $|p \sigma \rangle$ is independent of $n$ and $q$. 
Using this basis, 
we can expand the field operator $F(\vec r)$ on the tube surface as
\begin{equation}
F(\vec r) = \sum_{n,p,\sigma} 
\int \frac {dq}{2\pi} \psi_{nqp\sigma}(x,y) c_{np\sigma}(q)
= \sum_{n} F_n(x) \langle y| n \rangle,
\end{equation}
where the operator $c_{np\sigma}(q)$ destroys an electron in 
the state $|nqp\sigma \rangle$, and we introduce the 1D field operators
$F_n(x)$. Alternatively, using the basis of eigenstates of $H_0$, 
$F(\vec r)$ can be expanded as
\begin{equation}
F(\vec r) = \sum_{n,a,\sigma} 
\int \frac {dq}{2\pi} \psi_{nqa\sigma}(x,y) c_{na\sigma}(q) ,
\end{equation}
where the operators $c_{na\sigma}(q)$ destroy conduction ($a=+$)
or valence ($a=-$) electrons with spin $\sigma$ in subband $n$.
Notice that in what follows the spin index is left implicit.
The relation between the operators $c_{na}$ and $c_{np}$ is 
easily found to be
\begin{equation}
\left( \begin{array}{c} c_{n+}(q) \\ c_{n-}(q) \end{array} \right)=
\frac{1}{\sqrt{2}} \left( \begin{array}{cc} e^{- i\theta_n(q)/2} & 
e^{i\theta_n(q)/2}\\ e^{-i\theta_n(q)/2} & -e^{i\theta_n(q)/2}
\end{array} \right)
\left( \begin{array}{c} c_{nA}(q) \\ c_{nB}(q) \end{array}\right) .
\end{equation}

We now proceed by treating 
the spin-orbit hamiltonian using perturbation theory.
First, we diagonalize $H_0-\mu N$ for a fixed transverse subband $n$,
\begin{eqnarray*}
H^{(n)}_0 - \mu N^{(n)} &=& v \int dx \,
F^\dagger_n [k_\perp(n) \tau_1 + (-i\partial_x) \tau_2 - \mu ] F_n \\
&=& \sum_{a=\pm} \int \frac{dq}{2\pi} [a\epsilon_n(q) - \mu] c^\dagger_{na}
c^{}_{na}.
\end{eqnarray*}
Next we expand around the Fermi points $\pm q_n$ defined 
in Eq.~(\ref{vnqn}), which
introduces right- and left-movers, $r=\pm=R/L$, as the relevant
low-energy degrees of freedom.  For small deviations $k$ from $\pm q_n$,
Taylor expansion yields $\epsilon_n(\pm q_n +k) \simeq \mu \pm v_n k$, 
where $v_n$ is given in Eq.~(\ref{vnqn}).
Since we assumed $\mu>0$, we may now restrict ourselves to the 
conduction band,  $a=+$. For the hamiltonian, we then obtain
\begin{eqnarray*}
H^{(n)}_0 -\mu N^{(n)} & =  &
\sum_{r=\pm} v_n \int \frac{dk}{2\pi} ( r k) c^\dagger_{nr}(k)
c^{}_{nr}(k)\\ &=& \sum_{r=\pm} v_n \int dx \; \psi^\dagger_{nr} (-ir\partial_x) 
\psi^{}_{nr} ,
\end{eqnarray*}
where $c_{nr}(k)\equiv c_{n+}(rq_n+k)$ and 
$\psi_{nr}(x) = \int \frac{dk}{2\pi} e^{ikx} c_{nr}(k)$.
This introduces $R/L$-moving 1D fermion operators for each subband $n$
(and spin $\sigma$).
The relation of these 1D fermions with the original 
operator $F_n(x)$ is given by
\begin{eqnarray}\label{fn}
F_n(x) &=& e^{iq_n x} \int \frac{dk}{2\pi} \frac{e^{ikx}}{\sqrt{2}}
\left(\begin{array}{c} e^{i\theta_n(q_n)/2} \\ e^{-i\theta_n(q_n)/2} 
\end{array} \right) 
c_{nR}(k) \nonumber\\ 
& + & e^{- iq_n x} \int \frac{dk}{2\pi} \frac{e^{ikx}}{\sqrt{2}} 
\left(\begin{array}{c} e^{-i\theta_n(q_n)/2} \\ e^{i\theta_n(q_n)/2} 
\end{array} \right) c_{nL}(k) . 
\end{eqnarray}
Notice that, while in general the unitary 
transformation from sublattice space to the conduction/valence band
description depends on longitudinal momentum, in the continuum limit,
one can use the transformation directly at the Fermi momenta. 
This is consistent with the neglect of band curvature effects
implicit in the linearization 
of the dispersion relation, which  
is unproblematic away from van Hove singularities associated with
the onset of new subbands \cite{hugle}. At these points, the concept
of $R/L$-movers breaks down, and some of our conclusions below may change.

Next we express the Rashba hamiltonian (\ref{soham}) 
in terms of  $R/L$-movers. 
The first term results in
\begin{equation}\label{rasb}
H^{\parallel}_{so} = \frac{ u_\parallel v}{\mu}  \sum_{nr} 
k_\perp (n)
\int \frac{dk}{2\pi} 
 c^\dagger_{nr}(k) \sigma_3 c^{}_{nr}(k) .
\end{equation}
The presence of the factor $k_\perp(n)$ results from 
a careful treatment of the phases in Eq.~(\ref{fn}).
In Eq.~(\ref{rasb}) we omit an additional term  
mixing right- and left-movers. This term contains a 
rapidly oscillating factor $e^{\pm 2 i q_nx}$ and 
therefore is strongly suppressed by momentum conservation.  
The second term in Eq.~(\ref{soham}) 
again contains the oscillating phase factor
$e^{\pm i(q_n \pm q_{n+1})x}$, which leads to a drastic suppression of
$H^\perp_{so}$ at low energies and long wavelengths. 
Of course, this argument relies 
in an essential way on the smallness of the coupling $u_\perp$,
as one expands around the hamiltonian $H_0$.
We conclude that away from van Hove singularities, 
the only important Rashba term is given by $H^\parallel_{so}$
in Eq.~(\ref{rasb}).  This term has the appearance of a static 
homogeneous but {\sl channel-dependent magnetic field}. 

\section{Oscillatory TMR effects in nanotubes}\label{sec4}

In this section we will analyze the consequences of our findings
regarding spin-orbit couplings in CNTs, 
see Eqs.~(\ref{rasb}), 
for the observability of spin precession effects encoded in the
Datta-Das oscillations of the TMR.  Based on 
our expressions, it is possible to estimate the 
order of magnitude of this effect. 

For a concrete estimate, let us put $E_0=0.2eV_G/(\kappa D)$, 
where $D$ is the gate-tube distance, $V_G$ the gate voltage,
and $\kappa$ denotes the dielectric constant of the substrate. 
 For a given channel $n$, the Rashba-induced energy splitting is then
easily estimated as
\[
\frac{\Delta E_n}{eV_G}= 
(\gamma_1 + \gamma_2) \frac{0.6dv}{\mu} \frac{|n+n_0|}{R}
\frac{\alpha^2 \beta d^4\lambda_c^2}{4\kappa D},
\]
where $\lambda_c=\hbar/mc= 3.86\times 10^{-13}$~m is the Compton length.
Plugging in the definition of $\alpha,\beta$, we get
\begin{equation}
\label{splitting}
\frac{\Delta E_n}{eV_G} =\frac{0.6(\gamma_1+\gamma_2)}{256\pi \kappa} (d/a_0)^5
 \frac{\lambda_c^2}{D a_0} \frac{ |n+n_0| }{k_F R}.
\end{equation}
Bands with small $n$ are only  weakly split, and hence do
not contribute to oscillatory TMR behavior. 
This argument suggests that Datta-Das oscillations in
principle could survive in a CNT, even when there are 
many channels.  The major
contribution will come just from the few bands with the largest $n$.

To estimate the accumulated phase difference due to the different
precession length of the two split eigenstates, 
let us put $(n+n_0)/(k_F R)\to 1$, which represents the dominant contribution,
and set $\kappa=1$.  Then Eq.~(\ref{splitting}) 
gives as order-of-magnitude estimate
\begin{equation}
\Delta E/(eV_G)  \approx 2 \times 10^{-6} \ a_0/D.
\end{equation}
Even when assuming a very close-by gate, 
this gives only a tiny splitting, in retrospect justifying perturbation
theory.  This splitting now translates into a 
momentum splitting $\Delta k_n= \Delta E/v_n$,
and hence into a precession phase mismatch along the CNT of length 
$L$ \cite{datta}. For the $n$th band, this phase difference is
\begin{equation}\label{phase}
\Delta \phi_n= \Delta k_n L\approx 2 \times 10^{-6} 
\frac{L}{D} \frac{eV_G}{\hbar v_n/a_0}. 
\end{equation} 
This phase difference should be of order $2\pi$ to allow for 
the observation of Datta-Das oscillatory TMR effects \cite{datta}.

Away from a van Hove singularity, Eq.~(\ref{phase}) predicts 
that oscillations appear on a gate voltage scale of the order
of $10^6$ to $10^7$~V for $L\approx D$, which would make Datta-Das
oscillations unobservable. 
This argument also shows that this interpretation can be ruled
out for the parameters relevant for the Basel experiment \cite{sahoo}.
{}From Eq.~(\ref{phase}), 
we can then suggest several ways to improve the situation. 
First, one should use very long CNTs, while at the same time keeping
the gate very close, and second, an enhancement can be 
expected close to van Hove singularities. Of course, very close
to a van Hove singularity, some of our arguments above break down,
but the general tendency can nevertheless be read off from Eq.~(\ref{phase}).
Furthermore, electron-electron interactions
can also enhance spin-orbit effects \cite{hausler1,chenraikh}.  

To conclude, we have presented a detailed microscopic derivation
of Rashba spin-orbit coupling in carbon nanotubes.  It turns out that 
the Rashba SO coupling is small,
and therefore the prospects for observing spin-precession effects
like Datta-Das oscillations in the tunneling magnetoresistance
are not too favorable.  However, for very long CNTs, close-by gates,
and in the vicinity of a van Hove singularity, the requirements
for observability of these effects could be met in practice.

\ack
We thank T. Kontos and C. Sch\"onenberger for motivating this study.
Support by the DFG (Gerhard-Hess program) and by the EU (DIENOW network)  
is acknowledged.

\end{document}